Franziska Hummel, Marcus Tegel, Birgit Gerke, Rainer Pöttgen and Dirk Johrendt[*]

# Synthesis, Crystal Structure and Magnetism of $Eu_3Sc_2O_5Fe_2As_2$

**Abstract:** The iron arsenide $Eu_3Fe_2O_5Fe_2As_2$ was synthesized at 1173-1373 K in a resistance furnace and characterized by X-ray powder diffraction with Rietveld analysis: $Sr_3Fe_2O_5Cu_2S_2$-type, $I4/mmm$, $a$ = 406.40(1) pm, $c$ = 2646.9(1) pm. Layers of edge-sharing $FeAs_{4/4}$ tetrahedra are separated by perovskite-like oxide blocks. No structural transition occurs in the temperature range from 10 to 300 K. Magnetic measurements have revealed Curie-Weiss behavior with an effective magnetic moment of 7.79 $\mu_B$ per europium atom in agreement with the theoretical value of 7.94 $\mu_B$ for $Eu^{2+}$. A drop in the magnetic susceptibility at 5 K indicates possible antiferromagnetic ordering. [151]Eu and [57]Fe Mössbauer spectroscopic measurements have confirmed a beginning cooperative magnetic phenomenon by showing significantly broadened spectra at 4.8 K compared to those at 78 K.

**\*Corresponding author: Dirk Johrendt,** Department Chemie, Ludwig-Maximilians-Universität München, Butenandtstrasse 5-13, D-81377 München, Germany, E-mail: johrendt@lmu.de

**Franziska Hummel:** Department Chemie, Ludwig-Maximilians-Universität München, Butenandtstrasse 5-13, D-81377 München, Germany

**Marcus Tegel:** Fraunhofer-Institut für Fertigungstechnik und Angewandte Materialforschung IFAM, Winterbergstraße 28, D-01277 Dresden

**Birgit Gerke and Rainer Pöttgen:** Institut für Anorganische und Analytische Chemie, Universität Münster, Corrensstrasse 30, 48149 Münster, Germany

# 1 Introduction

Since the discovery of superconductivity in LaO$_{1-x}$F$_x$FeAs [1] (so-called 1111 type) many other structure types of iron-based superconductors have been identified [2-7]. Among these the 122 type [8], where layers of edge-sharing FeAs$_{4/4}$ tetrahedra are separated by a monolayer of ions, and the 32522 type [9], which exhibits larger perovskite-like interlayers between the FeAs layers. Despite the variety of compounds reported, only few europium compounds are known. Particularly EuFe$_2$As$_2$ has been intensively studied, as its properties are strongly related to those of the isostructural alkaline earth compounds. Similar to $AE$Fe$_2$As$_2$ ($AE$ = Ca, Sr, Ba), the parent compound undergoes a structural transition (at 190 K in the case of EuFe$_2$As$_2$ [10]), and superconductivity can be induced by suppressing the transition either by applying pressure [11] or by doping [12-18]. However, EuFe$_2$As$_2$ is a special case among the 122 parent compounds of iron-based superconductors due to the magnetism of europium. In addition to the antiferromagnetic stripe-like ordering of the iron moments at 190 K, it exhibits a second magnetic transition at 19 K where the europium moments order antiferromagnetically ($A$ type) [10]. Although the ordering of the Eu$^{2+}$ moments is not suppressed when superconductivity is induced, it seems to influence the superconducting properties. In EuFe$_2$As$_2$ [19] under pressure, and in Eu$_{1-x}$Sr$_x$Fe$_{2-y}$Co$_y$As$_2$ [20], a small but significant increase in resistivity has been observed at the ordering temperature of the europium moments.

Here we report the existence of the first europium compound of the 32522-type iron pnictides, where the FeAs layers are widely separated by perovskite-like oxide layers. Up to now, the only known iron pnictides with this structure are Sr$_3$Sc$_2$O$_5$Fe$_2$As$_2$ [21], Ba$_3$Sc$_2$O$_5$Fe$_2$As$_2$ [22], and Ca$_3$Al$_2$O$_{5-y}$Fe$_2$$Pn$$_2$ ($Pn$ = As, P) [23]. Only the calcium compounds show superconductivity, and by doping the strontium compound with titanium, traces of superconductivity have been observed in Sr$_3$Sc$_{2-x}$Ti$_x$O$_5$Fe$_2$As$_2$ [24].

# 2 Experimental

A polycrystalline sample of Eu$_3$Sc$_2$O$_5$Fe$_2$As$_2$ was synthesized by heating a stoichiometric mixture of Eu, Sc, FeO and As$_2$O$_3$. The reaction mixture was transferred into an alumina crucible and sealed in a silica ampoule under argon atmosphere. The sample was heated up to 1173 K for 20 h at rates of 50 and 200 K/h for heating and cooling, respectively. Afterwards, the sample was ground in an agate mortar, pressed into a pellet, and sintered for 60 h at 1373 K (heated at a rate of 100 K/h and cooled at 200 K/h). This sintering step was performed twice.

Temperature dependent X-ray powder-diffraction patterns between 300 and 10 K (10 K step size) were recorded using a Huber G670 Guinier Imaging Plate diffractometer (Co-K$\alpha_1$ radiation). The data were pre-processed with the program HConvert [25]. For Rietveld refinements of the data the TOPAS package [26] was used using the fundamental parameter approach for generating the reflection profiles. Spherical harmonics functions were used to describe the preferred orientation of the crystallites. Shape anisotropy effects were described by the approach of *Le Bail* and *Jouanneaux* [27]. Magnetic measurements were performed using a SQUID magnetometer (MPMS-XL5, Quantum Design, Inc.).

The 21.53 keV transition of $^{151}$Eu with an activity of 130 MBq (2 % of the total activity of a $^{151}$Sm:EuF$_3$ source) and a $^{57}$Co/Rh source were used for the Mössbauer spectroscopic characterizations. The measurements were conducted in transmission geometry with a commercial nitrogen-bath (78 K) and helium-flow (5 K) cryostat, while the sources were kept at room temperature. 100 ($^{151}$Eu measurements) and 40 mg ($^{57}$Fe measurements) of Eu$_3$Sc$_2$O$_5$Fe$_2$As$_2$ were placed in thin-walled PVC containers with optimized thicknesses of about 16.4 mg Eu cm$^{-2}$ and 1.6 mg Fe cm$^{-2}$. Fitting of the spectra was performed with the NORMOS-90 program system [28].

## 3 Results and discussion

### 3.1 Crystal chemistry

Eu$_3$Sc$_2$O$_5$Fe$_2$As$_2$ was obtained as a black polycrystalline air-stable sample. The crystal structure and the sample composition were analyzed via Rietveld refinements of the X-ray powder data at 300 K (Fig. 1): 91 % Eu$_3$Sc$_2$O$_5$Fe$_2$As$_2$, 7 % EuFe$_2$As$_2$, 1 % Sc$_2$O$_3$, and 1 % FeO (*wt*-%). Crystallographic data are compiled in Table 1. Further details of the structure determination may be obtained from: Fachinformationszentrum Karlsruhe, 76344 Eggenstein-Leopoldshafen, Germany, by quoting the Registry No. CSD-429704 (fax: (-49)7247-808-666; e-mail: crysdata@fiz-karlsruhe.de, http://www.fiz-karlsruhe.de/request_for_deposited_data.html).

$Eu_3Sc_2O_5Fe_2As_2$ crystallizes in the $Sr_3Fe_2O_5Cu_2S_2$-type structure [29] with lattice parameters $a$ = 406.40(1) pm, $c$ = 2646.9(1) pm, and consists of tetrahedral iron-arsenide layers separated by a perovskite-like separating oxide layer (Fig. 2). Eu1 is coordinated by 8+4 oxide ions similar to perovskites like $EuTiO_3$ [30], where all 12 Eu–O contacts are equal. Eu2 is eightfold coordinated by four oxide- and four arsenide neighbors. Scandium is surrounded by five oxide ions, which form a square pyramid. The Fe–As bonds (243.76(6) pm) and As-Fe-As angles (2 × 112.9(1)°, 4 × 107.8(1)°) in the $FeAs_{4/4}$ tetrahedra are close to the values found in related compounds like $Sr_3Sc_2O_5Fe_2As_2$ (Fe–As bonds: 243.56(9) pm, As-Fe-As angles: 2 × 113.3(1)°, 4 × 107.6(1)°) [9]. However, the tetrahedra are strongly compressed as compared to those in $EuFe_2As_2$ (Fe–As 256.45(5) pm, As-Fe-As 2 × 99.4(1)°, 4 × 114.7(1)°) [31] which becomes superconducting upon doping.

Temperature dependent X-ray data revealed no signs of a structural transition down to 10 K (Fig. 3). While the lattice parameter $a$ is contracted by only 0.2 % upon cooling, $c$ decreases by 0.6 % but the course of both lattice parameters as a function of temperature does not imply any anomaly which would indicate a structural transition. The reflections at $2\theta$ = 15, 25, and 29 ° appearing below 250 K are due to the sample environment and the reflection at about 42 °$2\theta$ belongs to the impurity phase $EuFe_2As_2$. This phase undergoes a structural transition below 190 K (tetragonal $I4/mmm$ to orthorhombic $Fmmm$ [32]).

### 3.2 Magnetic properties

Magnetic susceptibility measurements show Curie-Weiss behavior between 5 and 300 K (Fig. 4). The fit reveals an effective magnetic moment $\mu_{eff}$ = 13.49(1) $\mu_B$ per formula unit, equivalent to 7.79 $\mu_B$ per europium atom, close to the theoretical value of $Eu^{2+}$ of 7.94 $\mu_B$ [33]. The negative Weiss-constant $\theta$ = −9.5(3) K together with the drop in $\chi(T)$ near 5 K suggests antiferromagnetic ordering at low temperatures.

### 3.3 $^{57}$Fe and $^{151}$Eu Mössbauer spectroscopy

$^{151}$Eu and $^{57}$Fe spectra of $Eu_3Sc_2O_5Fe_2As_2$ are presented in Figs. 5 and 6 together with

transmission integral fits (Table 2). In accordance with the determined crystal structure, the $^{151}$Eu signal could be well reproduced by a superposition of two resonances with isomer shifts of –13.26(3) and –10.84(2) mm s$^{–1}$ indicating Eu$^{2+}$. The areas of both contributions were kept fixed with the ideal values derived from the multiplicity of the atomic Eu1 and Eu2 sites. With respect to its nearly ideal cuboctahedral coordination by 8+4 oxygen atoms, Eu1 was assumed to be the nucleus of less electron density (more ionic) compared to Eu2 and therefore the origin of the resonance at –13.26(3) mm s$^{–1}$. The isomer shifts of ionic Eu1 compare well with the values of the perovskites EuTiO$_3$ (–13.5 mm s$^{–1}$) and EuZrO$_3$ (–14.1 mm s$^{–1}$) [34, 35] as well as of several borate glasses containing divalent europium [36]. The Eu2 signal is close to the value observed for EuFe$_2$As$_2$ (–11.3 mm s$^{–1}$) [37].

Furthermore, a quadrupole splitting for Eu1 was excluded because of the symmetric arrangement around the nucleus (almost cubic site symmetry). A small resonance at an isomer shift of 0.80(4) mm s$^{–1}$ indicates trivalent europium possibly due to extrinsic Eu$^{3+}$ because of surface oxidation.

The $^{57}$Fe spectrum of Eu$_3$Sc$_2$O$_5$Fe$_2$As$_2$ shows a single signal with isomer shifts of 0.35(1) and 0.47(1) mms$^{–1}$ (intermetallic iron [32, 38, 39]) at ambient temperature and 78 K, respectively, and a small interaction between quadrupole and electric field gradient resulting in a splitting of 0.20(1) mm s$^{–1}$.

With regard to the suggested magnetic ordering, $^{151}$Eu and $^{57}$Fe Mössbauer spectroscopic measurements were repeated at 4.8 K (Figs. 5 and 6), quite close to the observed drop in the magnetic susceptibility data. Both the $^{151}$Eu and $^{57}$Fe spectra clearly show broadened resonances confirming a cooperative magnetic phenomenon. A fitting of the obtained spectra in order to determine the magnitude of the hyperfine field was difficult. On the one hand there are too many parameters that cannot be fitted confidentially; on the other hand at 4.8 K our experimental set up is at its limit. The temperature varies in the range of about ±0.3 K which has an influence on the exact splitting, especially for measurements close to the ordering temperature, leading to too many correlations within the refinement procedures.

## Summary

$Eu_3Sc_2O_5Fe_2As_2$ is a new member of the 32522-type iron pnictides with the $Sr_3Fe_2O_5Cu_2S_2$-type structure. Similar to the already known members of this group, we found no signs for a structural phase transition down to 10 K. Magnetic measurements revealed an effective magnetic moment of 7.79 $\mu_B$ per europium atom and possible antiferromagnetic ordering below 5 K. $^{151}$Eu and $^{57}$Fe Mössbauer spectroscopic experiments have confirmed the magnetic data by showing resonances for $Eu^{2+}$ and intermetallic iron. Low-temperature measurements showed a beginning magnetic ordering. As europium and alkaline earth compounds show a strong similarity for the 122-type iron pnictides, and traces of superconductivity were reported for $Sr_3Sc_{2-x}Ti_xO_5Fe_2As_2$, one could expect that $Eu_3Sc_2O_5Fe_2As_2$ might become superconducting upon doping.

**Table 1** Crystallographic data for $Eu_3Sc_2O_5Fe_2As_2$ at 300 K.

| | | |
|---|---|---|
| space group | $I4/mmm$ (no. 139) | |
| molar mass, g mol$^{-1}$ | 887.334 | |
| lattice parameter $a$, pm | 406.40(1) | |
| lattice parameter $c$, pm | 2646.9(1) | |
| cell volume, nm³ | 0.43717(2) | |
| density, g cm$^{-3}$ | 6.74(1) | |
| Z | 2 | |
| data points | 19001 | |
| reflections (all phases) | 161 | |
| refined parameters | 88 | |
| $R_P$, $wR_P$ | 0.017, 0.025 | |
| $R_{bragg}$, $\chi^2$ | 0.010, 1.094 | |
| atomic parameters | | |
| Eu1 | $2b$ (0, 0, 1/2) | $U_{iso}$ = 186(6) pm² |
| Eu2 | $4e$ (0, 0, $z$); $z$ = 0.3583(1) | $U_{iso}$ = 62(5) pm² |
| Sc | $4e$ (0, 0, $z$); $z$ = 0.0722(1) | $U_{iso}$ = 58(7) pm² |
| O1 | $8g$ (0, 1/2, $z$); $z$ = 0.0838(2) | $U_{iso}$ = 285(13) pm² |
| O2 | $2a$ (0, 0, 0) | $U_{iso}$ = 285(13) pm² |
| Fe | $4d$ (0, 1/2, 1/4) | $U_{iso}$ = 56(8) pm² |
| As | $4e$ (0, 0, $z$); $z$ = 0.1991(1) | $U_{iso}$ = 253(6) pm² |

**Table 2** Fitting parameters of $^{151}$Eu and $^{57}$Fe Mössbauer spectroscopic measurements at ambient temperature and at 78 K. $\delta$ = isomer shift, $\Delta E_Q$ = electric quadrupole splitting, $\Gamma$ = experimental line width. Parameters marked with an asterisk were kept fixed during the fitting procedure.

| Compound | $\delta$ (mm s$^{-1}$) | $\Delta E_Q$ (mm s$^{-1}$) | $\Gamma$ (mm s$^{-1}$) | ratio (%) |
|---|---|---|---|---|
| $^{151}$**Eu**$_3Sc_2O_5Fe_2As_2$ | | | | |
| 78 K | −13.26(3) | 0* | 3.11(6) | 28* |
| | −10.84(2) | −3.3(2) | 2.8(1) | 57* |
| | 0.80(4) | 0* | 2.8(1) | 15(1) |
| $Eu_3Sc_2O_5$$^{57}$**Fe**$_2As_2$ | | | | |
| 298 K | 0.35(1) | 0.20(1) | 0.31(1) | 100 |
| 78 K | 0.47(1) | 0.20(1) | 0.32(1) | 100 |

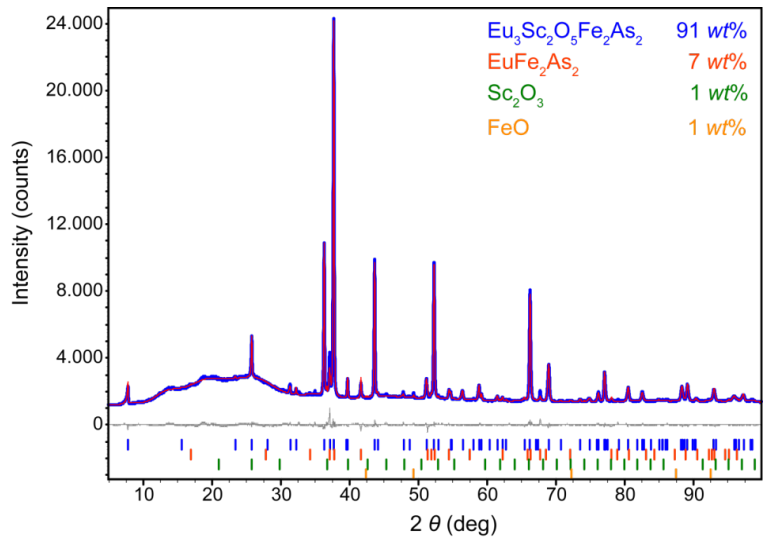

**Fig. 1** X-ray diffraction pattern at 300 K (blue) and Rietveld refinement (red) of $Eu_3Sc_2O_5Fe_2As_2$.

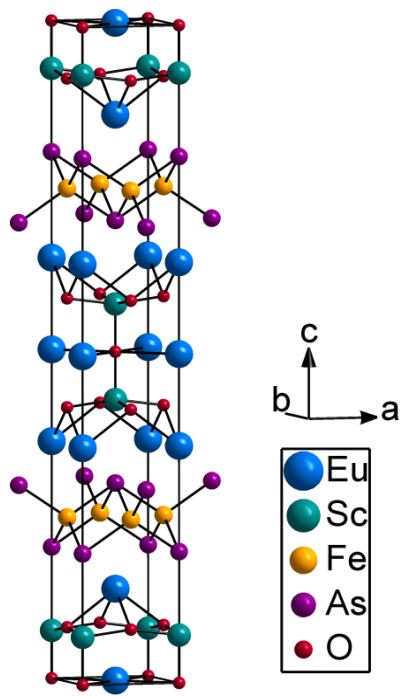

**Fig. 2** Crystal structure of $Eu_3Sc_2O_5Fe_2As_2$.

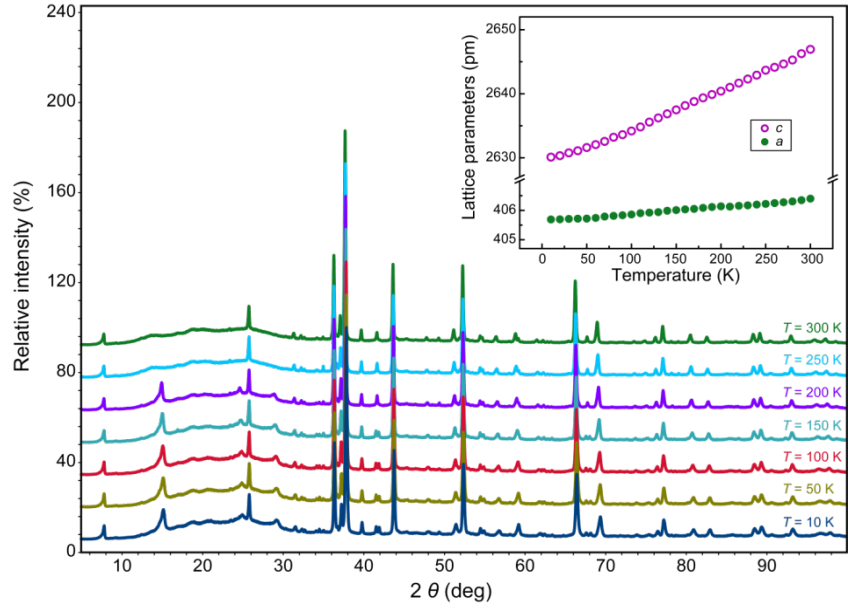

**Fig. 3** X-ray diffraction patterns of $Eu_3Sc_2O_5Fe_2As_2$ at different temperatures. Inset: Refined lattice parameters as a function of temperature.

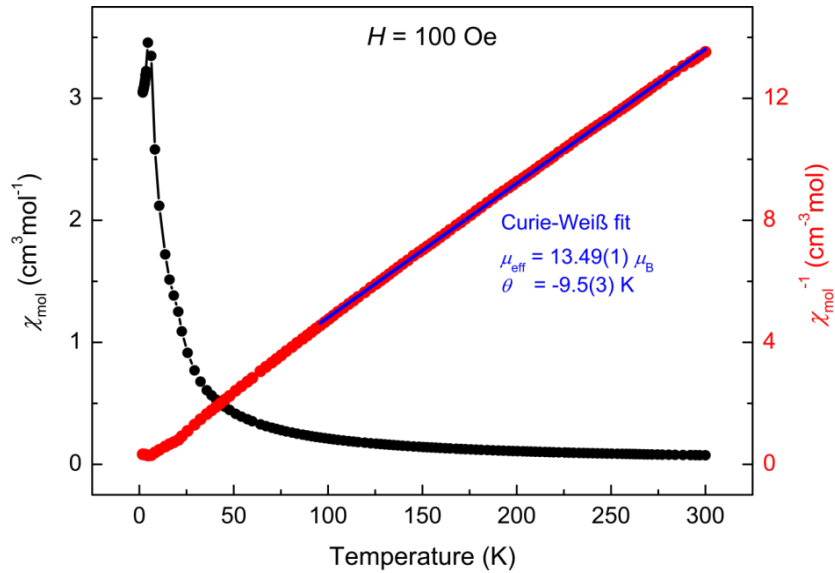

**Fig. 4** Magnetic susceptibility (black) and inverse magnetic susceptibility (red) of $Eu_3Sc_2O_5Fe_2As_2$ measured at 100 Oe. The Curie-Weiss fit is shown as blue line.

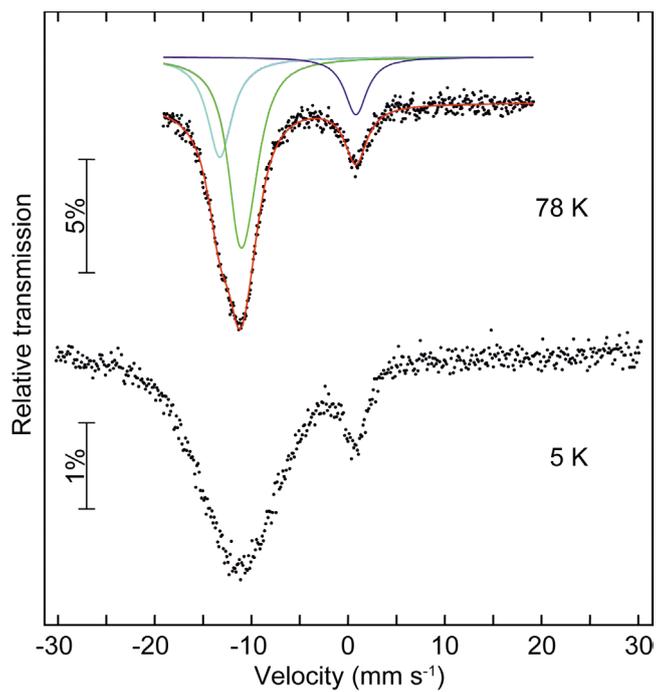

**Fig. 5** Experimental and simulated $^{151}$Eu Mössbauer spectra of $Eu_3Sc_2O_5Fe_2As_2$ at 78 and 5 K.

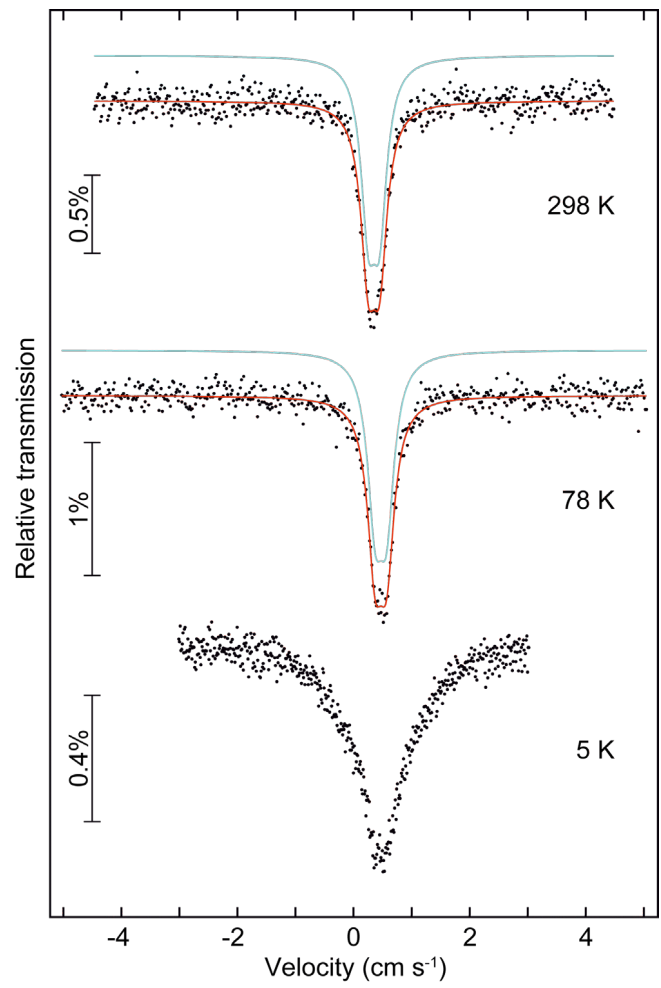

**Fig. 6** Experimental and simulated $^{57}$Fe Mössbauer spectra of $Eu_3Sc_2O_5Fe_2As_2$ at ambient temperature, 78 and 5 K.

**Synopsis:** Eu$_3$Fe$_2$O$_5$Fe$_2$As$_2$ contains layers of edge-sharing FeAs$_{4/4}$ tetrahedra that are separated by perovskite-like oxide blocks.

**TOC figure**

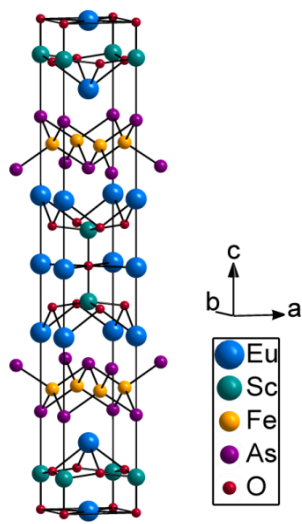